\documentclass[showpacs,preprintnumbers,twocolumn,
amsmath,amssymb,groupedaddress,superscriptaddress]{revtex4}

\usepackage{graphicx}
\usepackage{dcolumn}
\usepackage{bm}

\begin{document}

\title{Ground-state properties and symmetry energy of
neutron-rich and neutron-deficient Mg isotopes}

\author{M. K. Gaidarov}
\affiliation{Institute for Nuclear Research and Nuclear Energy,
Bulgarian Academy of Sciences, Sofia 1784, Bulgaria}

\author{P. Sarriguren}
\affiliation{Instituto de Estructura de la Materia, IEM-CSIC,
Serrano 123, E-28006 Madrid, Spain}

\author{A. N. Antonov}
\affiliation{Institute for Nuclear Research and Nuclear Energy,
Bulgarian Academy of Sciences, Sofia 1784, Bulgaria}

\author{E. Moya de Guerra}
\affiliation{Grupo de F\'{i}sica Nuclear, Departamento de
F\'{i}sica At\'{o}mica, Molecular y Nuclear,\\ Facultad de
Ciencias F\'{i}sicas, Unidad Asociada UCM-CSIC(IEM), Universidad
Complutense de Madrid, E-28040 Madrid, Spain}


\begin{abstract}
A comprehensive study of various ground-state properties of
neutron-rich and neutron-deficient Mg isotopes with $A$=20--36 is
performed in the framework of the self-consistent deformed
Skyrme-Hartree-Fock plus BCS method. The correlation between the
skin thickness and the characteristics related with the density
dependence of the nuclear symmetry energy is investigated for this
isotopic chain following the theoretical approach based on the
coherent density fluctuation model and using the Brueckner
energy-density functional. The results of the calculations show
that the behavior of the nuclear charge radii and the nuclear
symmetry energy in the Mg isotopic chain is closely related to the
nuclear deformation. We also study, within our theoretical scheme,
the emergence of an "island of inversion" at neutron-rich
$^{32}$Mg nucleus, that was recently proposed from the analyses of
spectroscopic measurements of $^{32}$Mg low-lying energy spectrum
and the charge rms radii of all magnesium isotopes in the $sd$
shell.
\end{abstract}


\maketitle

\section{Introduction}
The study of nuclear structure has advanced on the basis of the
shell structure associated with the magic numbers. This study,
however, has been carried out predominantly for stable nuclei,
which are on or near the $\beta$-stability line in the nuclear
chart and have been explored experimentally. Advances in
measurements of unstable nuclei have provided information on
exotic nuclei toward the neutron and proton drip lines. Close to
them, a large variety of formerly unknown nuclear configurations
has been observed. The magic numbers in such exotic systems can be
a quite intriguing issue. New magic numbers appear and some others
disappear in moving from stable to exotic nuclei in a rather novel
manner due to specific components of the nucleon-nucleon interaction
(see, for example, Ref.~\cite{Otsuka2001}).

Low-lying states of neutron-rich nuclei around the neutron number
$N$=20 attract a great interest, as the spherical configurations
associated with the magic number disappear in the ground states.
For $^{32}$Mg, from the observed population of the excited
$0^{+}_{2}$ state (found at 1.058 MeV) in the $(t,p)$ reaction on
$^{30}$Mg, it is suggested \cite{Wimmer2010} that the $0^{+}_{2}$
state is a spherical one coexisting with the deformed ground state
and that their relative energies are inverted at $N$=20. Very
recently, a new signature of an existence of "island of inversion"
\cite{Warburton90} has been experimentally tested by measuring the
charge radii of all magnesium isotopes in the $sd$ shell at
ISOLDE-CERN \cite{Yordanov2012} showing that the borderline of
this island lies between $^{30}$Mg and $^{31}$Mg. The new mass
measurements of the exotic nuclides $^{30-34}$Mg have led to a
value of 1.10(3) MeV for the empirical shell gap for $^{32}$Mg,
which reveals the lowest observed strength of nuclear shell
closure for a nuclide with a conventional magic number
\cite{Chaudhuri2013}. Previously, the Coulomb excitation studies
of $^{32}$Mg also confirmed the large deformation and pointed out
the vanishing of the $N$=20 shell gap \cite{Motobayashi95}. The
concept of "island of inversion" has been explored, for instance,
experimentally also to isotones with $N$=19 \cite{Christian2012}
and complementary information about its extent to at least $N$=22
for the Ne isotopes was obtained \cite{Doornenbal2009}.

On the theoretical side, the properties of neutron-rich Mg
isotopes have been mainly studied in the framework of the Shell
Model \cite{Warburton90,ShellModel}, as well as within
self-consistent mean-field approaches, namely, the Skyrme
Hartree-Fock (HF) method with pairing correlations (see, for
example,
Refs.~\cite{Lenske98,Siiskonen99,Horiuchi2012,Horiuchi2014}) or
the Hartree-Fock-Bogoliubov (HFB) method \cite{Terasaki97}. The
ground-state properties of Mg isotopes from the proton-drip-line
nucleus $^{20}$Mg to the neutron-drip-line nucleus $^{40}$Mg have
been calculated using the nonlinear relativistic mean-field (RMF)
model with force parameters NL-SH \cite{Ren96}. The latter
predicts well the monotonic increase of the neutron skin thickness
with neutron excess, and also provides a good description of the
binding energies and the charge rms radii of Mg isotopes. The RMF
approach with the same NL-SH parameter set has been applied to
calculate the energy levels, the proton occupation probabilities
of the $1d_{5/2}$, $2s_{1/2}$, and $1d_{3/2}$ states for the
even-even nuclei $^{26-40}$Mg, as well as their charge density
distributions \cite{Wang2013}. It was shown that the level
inversion of $2s_{1/2}$ and $1d_{3/2}$ states that may occur for
the magnesium isotopes with more neutrons away from the stability
line can lead to a large measurable central charge depletion to
the charge density distributions for the neutron-rich Mg isotopes.
Recently, the constrained HFB plus local quasiparticle RPA method
has been used to describe a variety of quadrupole collective
phenomena in a unified way, particularly the shape fluctuations in
the ground and excited 0$^{+}$ states of $^{30}$Mg and $^{32}$Mg
nuclei \cite{Hinohara2011}. The quadrupole deformation properties
of the ground and low-lying excited states of the even-even
magnesium isotopes with $N$ ranging from 8 to 28 were successfully
described in Ref.~\cite{Guzman2002}, where an angular momentum
projected generator coordinate method with the Gogny force was
explored. The fully-microscopic antisymmetrized molecular dynamics
calculations with the same Gogny-D1S effective interaction have
well reproduced measured ground-state properties (spin-parity,
total binding energy, one-neutron separation energy, and matter
radii) of Mg isotopes \cite{Watanabe2014}. In
Ref.~\cite{Caurier2014}, the $N$=20 and $N$=28 "islands of
inversion" have been shown to merge in the magnesium chain,
enclosing all the isotopes between $N$=19 and $N$=30, by
performing large-scale shell-model calculations for their
description.

The nuclear deformation plays an important role to determine the
charge and matter radius of the neutron-rich nuclei, in particular
for the description of the structure of the Mg isotopes
\cite{Horiuchi2012,Terasaki97,Ren96}. A direct link between the
inversion of states in the deformed shell model and the change in
the quadrupole deformation takes place
\cite{Hamamoto2007,Yordanov2010}. Nuclear radii carry also
important information about shell effects and residual
interactions. The recent development of the radioactive ion beam
(RIB) facilities has made it possible to search for anomalous
structure effects, such as the halo and skin, that are related
with the isotope and isospin dependencies of the radii. Suzuki
{\it et al.} \cite{Suzuki98} reported experimental effective rms
matter radii of some Na and Mg nuclei. They concluded that the
increase of the rms matter radius is mainly due to an increase of
the rms neutron radius. The presence of a neutron skin was
predicted in heavy Na and Mg nuclei.

The systematic investigations of the nuclear size properties
provide an important information about the saturation property of
atomic nuclei. In particular, the neutron-skin thickness as a good
isovector indicator \cite{Reinhard2010} has attracted much
interest. In Ref.~\cite{Sarriguren2007} the formation of a neutron
skin, which manifests itself in an excess of neutrons at distances
larger than the radius of the proton distribution, was analyzed in
terms of various definitions. The obtained results are
illustrative for the range of the skin sizes to be expected
depending on the adopted skin definition. The knowledge of skin
thickness gives more insight into the properties of neutron-rich
nuclei and allows to resolve some of the basic features of the
equation of state (EOS) of asymmetric nuclear matter (ANM). The
neutron skin thickness is strongly correlated with the density
dependence of the neutron-matter EOS \cite{Brown2000,Typel2001}
and of the symmetry energy of nuclear matter (see, e.g., the
articles in the topical issue \cite{Li2014}). The precise
knowledge of these relations is essential for predicting the
structure of neutron stars, particularly their radii
\cite{Steiner2010}.

In our recent works we have studied the relation between the
neutron skin thickness and some nuclear matter properties in
finite nuclei, such as the symmetry energy at the saturation point
$s$, symmetry pressure $p_{0}$ (proportional to the slope of the
bulk symmetry energy), and asymmetric compressibility $\Delta K$,
for chains of medium-heavy and heavy spherical Ni ($A$=74--84), Sn
($A$=124--152), and Pb ($A$=202--214) nuclei \cite{Gaidarov2011}
and deformed Kr ($A$=82--120) and Sm ($A$=140--156) isotopes
\cite{Gaidarov2012}. Most of these nuclei are far from the
stability line and are of interest for future measurements with
radioactive exotic beams. For this purpose, a theoretical approach
to the nuclear many-body problem combining the deformed HF+BCS
method with Skyrme-type density-dependent effective interactions
\cite{vautherin} and the coherent density fluctuation model (CDFM)
\cite{Ant80,AHP} has been used to study nuclear properties of
finite nuclei. The analysis of the latter has been carried out on
the basis of the Brueckner energy-density functional (EDF) for
infinite nuclear matter \cite{Brueckner68,Brueckner69}. It has to
be mentioned that the used microscopic theoretical approach is
capable of predicting important nuclear matter quantities in
neutron-rich exotic nuclei and their relation to surface
properties of these nuclei, that is confirmed by the good
agreement achieved with other theoretical predictions and some
experimentally extracted ground-state properties.

In the light of the new precise spectroscopic measurements of the
neutron-rich $^{32}$Mg nucleus, which lies in the much explored
"island of inversion" at $N$=20, in the present work we aim to
perform a systematic study of the nuclear ground-state properties
of neutron-rich and neutron-deficient Mg isotopes with $A$=20--36,
such as charge and matter rms radii, two-neutron separation
energies, neutron, proton, and charge density distributions,
neutron (proton) rms radii and related with them thickness of the
neutron (proton) skins. The new data for the charge rms radii
\cite{Yordanov2012} is a challenging issue to test the
applicability of the mean-field description to light nuclei, thus
expecting to understand in more details the nuclear structure
revealed by them. The need of information for the symmetry energy
in finite nuclei, even theoretically obtained, is a major issue
because it allows one to constrain the bulk and surface properties
of the nuclear EDFs quite effectively. Therefore, following our
recent works \cite{Gaidarov2011,Gaidarov2012} we analyze the
correlation between the skin thickness and the characteristics
related to the density dependence of the nuclear symmetry energy
for the same Mg isotopic chain. Such an analysis may probe the
accurate account for the effects of interactions in our method
within the considered Mg chain, where the breakdown of the shell
model could be revealed also by the nuclear symmetry energy
changes. A special attention is paid to the neutron-rich $^{32}$Mg
nucleus by performing additional calculations modifying the
spin-orbit strength of the effective interaction, to check
theoretically the possible appearance of the "island of inversion"
at $N$=20.

The structure of this article is the following. In Sec.~II we
present briefly the deformed HF+BCS method formalism together with
the basic expressions for the considered ground-state properties,
as well as the CDFM formalism that provides a way to calculate the
symmetry energy in finite nuclei. The numerical results and
discussions are presented in Sec.~III. We draw the main
conclusions of this study in Sec.~IV.

\section{A brief summary of the formalism}

The results of the present work have been obtained from
self-consistent deformed Hartree-Fock calculations with density
dependent Skyrme interactions \cite{vautherin} and pairing
correlations. Pairing between like nucleons has been included by
solving the BCS equations at each iteration with a fixed pairing
strength that reproduces the  odd-even experimental
mass differences \cite{Audi2003}). Here we give briefly the basic
expressions of the nuclear ground-state properties examined for Mg
isotopes, as well as the CDFM scheme to calculate the nuclear
symmetry energy in finite nuclei.

We consider in this paper the Skyrme force SLy4 \cite{sly4}. We
also show results obtained from other parametrizations, namely Sk3
\cite{sk3} and  SGII \cite{sg2} because they are among the most
extensively used Skyrme forces and are  considered as standard
references.

The spin-independent proton and neutron densities are given by
\cite{Sarriguren2007,Guerra91}
\begin{equation}
\rho({\vec R})=\rho (r,z)=\sum _{i} 2v_i^2\rho_i(r,z)\, ,
\label{eq:21}
\end{equation}
where $r$ and $z$ are the cylindrical coordinates of ${\vec R}$,
$v_i^2$ are the occupation probabilities resulting from the BCS
equations and $\rho_i$ are the single-particle densities
\begin{equation}
\rho_i({\vec R})=  \rho_i(r,z)=|\Phi^+_i(r,z)|^2+
|\Phi^-_i(r,z)|^2 \label{eq:22}
\end{equation}
with
\begin{eqnarray}
\Phi^\pm _i(r,z)&=&{1\over \sqrt{2\pi}}\nonumber \\
&\times & \sum_{\alpha}\, \delta_{\Sigma, \pm 1/2}\,
\delta_{\Lambda,\Lambda^\mp}\, C_\alpha ^i\, \psi_{n_r}^\Lambda
(r) \, \psi_{n_z}(z) \label{eq:23}
\end{eqnarray}
and $\alpha=\{n_r,n_z,\Lambda,\Sigma\}$. In (\ref{eq:23}) the
functions $\psi^{\Lambda}_{n_r}(r)$ and $\psi_{n_z}(z)$ are
expressed by Laguerre and Hermite polynomials:
\begin{equation}
\psi^{\Lambda}_{n_r}(r)=\sqrt{\frac{n_r}{(n_r+\Lambda )!}} \,
\beta_{\perp}\, \sqrt{2}\, \eta^{\Lambda/2}\, e^{-\eta/2}\,
L_{n_r}^{\Lambda}(\eta) \, , \label{eq:24}
\end{equation}
\begin{equation}
\psi_{n_z}(z)= \sqrt{\frac{1}{\sqrt{\pi}2^{n_z}n_z!}} \,
\beta^{1/2}_z\, e^{-{\xi}^2/2}\, H_{n_z}(\xi) \label{eq:25}
\end{equation}
with
\begin{eqnarray}
\beta_z=(m\omega_z/\hbar )^{1/2}&,&\quad
\beta_\perp=(m\omega_\perp/\hbar )^{1/2},\nonumber \\
\quad \xi=z\beta_z&,&\quad \eta=r^2\beta_\perp ^2 \, .
\label{eq:26}
\end{eqnarray}

The multipole decomposition of the density can be written as
\cite{vautherin,Guerra91}
\begin{eqnarray}
\rho(r,z)& = &\sum_{\lambda} \rho_{\lambda}(R)
P_{\lambda}(\cos\theta)\nonumber \\
&=& \rho_0(R) + \rho_2(R)\, P_2(\cos \theta) + \ldots \, ,
\label{rhomult}
\end{eqnarray}
with multipole components $\lambda$
\begin{equation}
\rho_{\lambda}(R)=\frac{2\lambda +1}{2} \int_{-1}^{+1}
P_{\lambda}(\cos\theta) \rho(R\cos\theta,R\sin\theta)
d(\cos\theta) \, ,
\end{equation}
and normalization given by
\begin{equation}
\int \rho({\vec R}) d{\vec R} = X ;\qquad  4\pi \int
R^2dR\rho_0(R) = X \, ,
\end{equation}
with $X=Z,\, N$ for protons and neutrons, respectively.

The mean square radii for protons and neutrons are defined as
\begin{equation}
\langle r_{p,n}^2\rangle =\frac{ \int R^2\rho_{p,n}({\vec
R})d{\vec R}} {\int \rho_{p,n}({\vec R})d{\vec R}} \, ,
\label{r2pn}
\end{equation}
and the rms radii for protons and neutrons are given by
\begin{equation}
r_{p,n}=\langle r_{p,n}^2 \rangle ^{1/2} \, .
\label{rmsrnp}
\end{equation}
The matter rms radius can be obtained by
\begin{equation}
r_{m}=\sqrt{\frac{N}{A}r_{n}^{2}+\frac{Z}{A}r_{p}^{2}} \,\, ,
\label{rmsrm}
\end{equation}
where $A=Z+N$ is the mass number. Having the neutron and proton
rms radii [Eq.~(\ref{rmsrnp})], the neutron skin thickness is
usually estimated as their difference:
\begin{equation}
\Delta R=\langle r_{n}^2\rangle ^{1/2}-\langle r_{p}^2\rangle
^{1/2}.
\label{eq:28}
\end{equation}

The mean square radius of the charge distribution in a nucleus can
be expressed as
\begin{equation}
\langle r^2_{ch}\rangle = \langle r^2_{p}\rangle + \langle
r^2_{ch}\rangle _{p} + (N/Z) \langle r^2_{ch}\rangle _{n} +
r^2_{CM} + r^2_{SO} \, ,
\label{rch}
\end{equation}
where $\langle r^2_{p}\rangle$ is the mean square radius of the
point proton distribution in the nucleus (\ref{r2pn}), $\langle
r^2_{ch}\rangle_{p}$ and $\langle r^2_{ch}\rangle_{n}$ are the
mean square charge radii of the charge distributions in a proton
and a neutron, respectively. $r^2_{CM}$ is a small correction due
to the center of mass motion, which is evaluated assuming
harmonic-oscillator wave functions. The last term $r^2_{SO}$ is a
tiny spin-orbit contribution to the charge density.
Correspondingly, we define the charge rms radius
\begin{equation}
r_{\rm c}=\langle r_{ch}^2\rangle ^{1/2} \, .
\label{rmsrc}
\end{equation}

In the present work we calculate the symmetry energy $s(\rho)$ of
Mg isotopes on the basis of the corresponding definition for ANM.
The quantity $s^{ANM}(\rho)$, which refers to the infinite system
and therefore neglects surface effects, is related to the second
derivative of the energy per particle $E(\rho,\delta)$ using its
Taylor series expansion in terms of the isospin asymmetry
$\delta=(\rho_{n}-\rho_{p})/\rho$, where $\rho$, $\rho_{n}$ and
$\rho_{p}$ are the baryon, neutron and proton densities,
respectively, (see, e.g.,
\cite{Gaidarov2011,Gaidarov2012,Diep2003,Chen2011}):
\begin{eqnarray}
s^{ANM}(\rho)&=&\frac{1}{2}\left.
\frac{\partial^{2}E(\rho,\delta)}{\partial\delta^{2}} \right
|_{\delta=0}=
a_{4}+\frac{p_{0}^{ANM}}{\rho_{0}^{2}}(\rho-\rho_{0}) \nonumber \\
&+& \frac{\Delta K^{ANM}}{18\rho_{0}^{2}}(\rho-\rho_{0})^{2}+
\cdot\cdot\cdot \;.
\label{eq:1}
\end{eqnarray}
In Eq.~(\ref{eq:1}) the parameter $a_{4}$ is the symmetry energy
at equilibrium ($\rho=\rho_{0}$). In ANM the pressure
$p_{0}^{ANM}$
\begin{equation}
p_{0}^{ANM}=\rho_{0}^{2}\left.
\frac{\partial{s^{ANM}(\rho)}}{\partial{\rho}} \right
|_{\rho=\rho_{0}},
\label{eq:2}
\end{equation}
of the nuclear symmetry energy at $\rho_{0}$ governs its density
dependence and thus provides important information on the
properties of the nuclear symmetry energy at both high and low
densities. The slope parameter $L^{ANM}$ is related to the
pressure $p_{0}^{ANM}$ [Eq.~(\ref{eq:2})] by
\begin{equation}
L^{ANM}=\frac{3p_{0}^{ANM}}{\rho_{0}}.
\label{eq:3}
\end{equation}

In Refs.~\cite{Gaidarov2011,Gaidarov2012} we calculated the
symmetry energy, the pressure and the curvature for {\it finite
nuclei} applying the coherent density fluctuation model
\cite{Ant80,AHP}. The key ingredient element of the calculations
is the weight function that in the case of monotonically
decreasing local densities ($d\rho(r)/dr\leq 0$) can be obtained
using a known density distribution for a given nucleus:
\begin{equation}
|f(x)|^{2}=-\frac{1}{\rho_{0}(x)} \left. \frac{d\rho(r)}{dr}\right
|_{r=x},
\label{eq:9}
\end{equation}
where $\rho_{0}(x)=3A/4\pi x^{3}$ and with the normalization
$\int_{0}^{\infty}dx |f(x)|^{2}=1$. In the calculations, for the
density distribution $\rho(r)$ needed to obtain the weight
function $|f(x)|^{2}$ [Eq.~(\ref{eq:9})] we use the monopole term
$\rho_{0}(R)$ in the expansion (\ref{rhomult}).

It can be shown in the CDFM that under some approximation the
properties of {\it finite nuclei} can be calculated using the
corresponding ones for nuclear matter, folding them with the
weight function $|f(x)|^{2}$. Along this line, in the CDFM the
symmetry energy for finite nuclei and related with it pressure are
obtained as infinite superpositions of the corresponding ANM
quantities weighted by $|f(x)|^{2}$:
\begin{equation}
s=\int_{0}^{\infty}dx|f(x)|^{2}s^{ANM}(x),
\label{eq:10}
\end{equation}
\begin{equation}
p_{0}=\int_{0}^{\infty}dx|f(x)|^{2}p_{0}^{ANM}(x).
\label{eq:11}
\end{equation}
The explicit forms of the ANM quantities $s^{ANM}(x)$ and
$p_{0}^{ANM}(x)$ in Eqs.~(\ref{eq:10}) and (\ref{eq:11}) are
defined below. They have to be determined within a chosen method
for the description of the ANM characteristics. In the present
work, as well as in Refs.~\cite{Gaidarov2011,Gaidarov2012},
considering the pieces of nuclear matter with density
$\rho_{0}(x)$, we use for the matrix element $V(x)$ of the nuclear
Hamiltonian the corresponding ANM energy from the method of
Brueckner {\it et al.} \cite{Brueckner68,Brueckner69}:
\begin{equation}
V(x)=A V_{0}(x)+V_{C}-V_{CO},
\label{eq:13}
\end{equation}
where
\begin{eqnarray}
V_{0}(x)&=&37.53[(1+\delta)^{5/3}+(1-\delta)^{5/3}]\rho_{0}^{2/3}(x)\nonumber \\
&+&b_{1}\rho_{0}(x)+b_{2}\rho_{0}^{4/3}(x)+b_{3}\rho_{0}^{5/3}(x)\nonumber \\
&+&\delta^{2}[b_{4}\rho_{0}(x)+b_{5}\rho_{0}^{4/3}(x)+b_{6}\rho_{0}^{5/3}(x)]
\label{eq:14}
\end{eqnarray}
with
\begin{eqnarray}
b_{1}&=&-741.28, \;\;\; b_{2}=1179.89, \;\;\; b_{3}=-467.54,\nonumber \\
b_{4}&=&148.26, \;\;\;\;\;\; b_{5}=372.84, \;\;\;\; b_{6}=-769.57.
\label{eq:15}
\end{eqnarray}
In Eq.~(\ref{eq:13}) $V_{0}(x)$ is the energy per particle in
nuclear matter (in MeV) accounting for the neutron-proton
asymmetry, $V_{C}$ is the Coulomb energy of protons in a flucton,
and $V_{CO}$ is the Coulomb exchange energy. Thus, using the
Brueckner method, the symmetry energy $s^{ANM}(x)$ and the
pressure $p_{0}^{ANM}(x)$ for ANM with density $\rho_{0}(x)$ have
the forms:
\begin{eqnarray}
s^{ANM}(x)&=&41.7\rho_{0}^{2/3}(x)+b_{4}\rho_{0}(x) \nonumber \\
&+&b_{5}\rho_{0}^{4/3}(x)+b_{6}\rho_{0}^{5/3}(x),
\label{eq:18}
\end{eqnarray}
and
\begin{eqnarray}
p_{0}^{ANM}(x)&=&27.8\rho_{0}^{5/3}(x)+b_{4}\rho_{0}^{2}(x) \nonumber \\
&+&\frac{4}{3}b_{5}\rho_{0}^{7/3}(x)+\frac{5}{3}b_{6}\rho_{0}^{8/3}(x).
\label{eq:19}
\end{eqnarray}
In our approach (see also \cite{Gaidarov2011,Gaidarov2012})
Eqs.~(\ref{eq:18}) and (\ref{eq:19}) are used to calculate the
corresponding quantities in finite nuclei $s$ and $p_{0}$ from
Eqs.~(\ref{eq:10}) and (\ref{eq:11}), respectively. We note that
in the limit case when $\rho(r)=\rho_{0}\Theta (R-r)$ and
$|f(x)|^{2}$ becomes a $\delta$ function [see Eq.~(\ref{eq:9})],
Eq.~(\ref{eq:10}) reduces to $s^{ANM}(\rho_{0})=a_{4}$.

\section{Results and discussion}
\subsection{Ground-state properties}

\begin{figure}
\centering
\includegraphics[width=78mm]{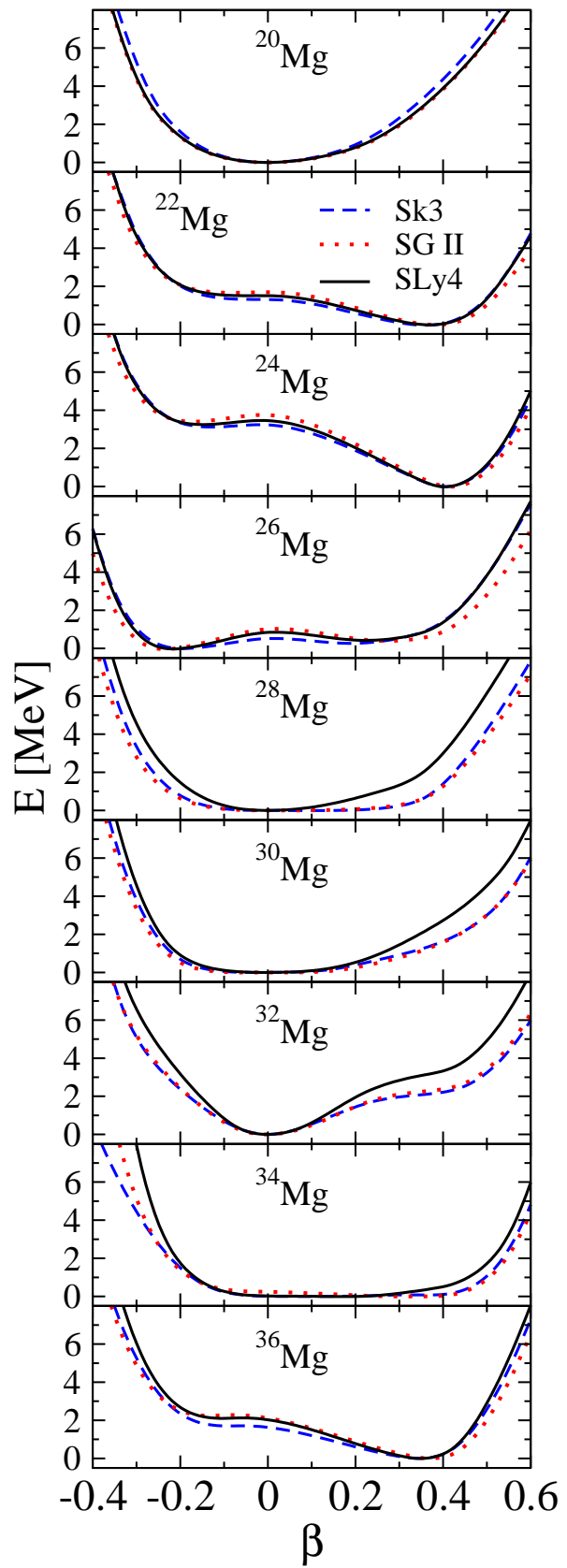}
\caption[]{(Color online) Potential energy curves in even-even Mg
isotopes ($A$=20--36) obtained from HF+BCS calculations with three
different Skyrme forces.
\label{fig1}}
\end{figure}

We start our analysis by showing the variation of the binding
energy $E$ of all Mg isotopes with $A$=20--36 as a function
of their quadrupole parameter $\beta$:
\begin{equation}
\beta=\sqrt{\frac{\pi}{5}}\frac{Q}{A\langle r^2\rangle ^{1/2}}\, ,
\label{eq:29}
\end{equation}
where $Q$ is the mass quadrupole moment and $\langle r^2\rangle
^{1/2}$ is the nucleus rms radius. The corresponding potential
energy curves obtained with three different Skyrme forces are
given in Fig.~\ref{fig1} for the even-even isotopes. Similar
profiles are obtained by using constant pairing gap parameters in
the BCS calculations.

\begin{figure}
\centering
\includegraphics[width=80mm]{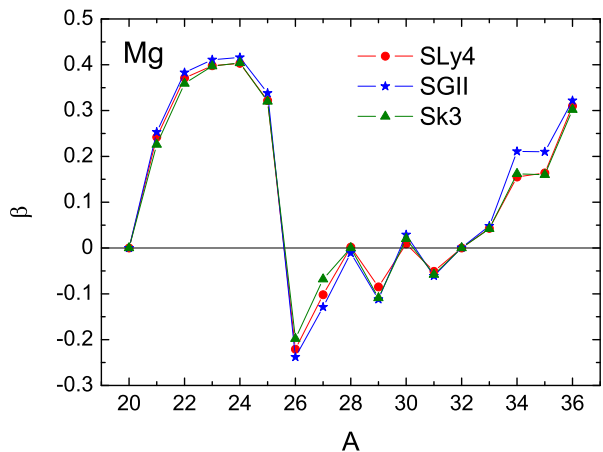}
\caption[]{(Color online) The quadrupole parameter $\beta$ of
the ground state as a function of the mass number $A$ for Mg isotopes
($A$=20--36) in the cases of SLy4, SGII, and Sk3 forces.
\label{fig2}}
\end{figure}

The study of the shape evolution in neutron-rich and
neutron-deficient Mg isotopes is further extended by presenting in
Fig.~\ref{fig2} the evolution of the quadrupole parameter $\beta$
of the ground states as a function of the mass number $A$. A
combined analysis of both Fig.~\ref{fig1} and Fig.~\ref{fig2}
shows that, as expected, the semi-magic $^{20}$Mg isotope ($N$=8)
is spherical. As the number of neutrons increases we start
populating first the $d_{5/2}$ orbital between  $^{22}$Mg and
$^{26}$Mg, leading to prolate and oblate defomations in the energy
profiles. Prolate shapes with $\beta \sim 0.4$ are ground states
in  $^{22}$Mg and  $^{24}$Mg, whereas an oblate shape at around
$\beta \sim -0.25$ is developed in $^{26}$Mg in competition with a
prolate shape at around $\beta \sim 0.35$. Due to the conjunction
of the $N$=$Z$=12 deformed shell effects, the nucleus $^{24}$Mg is
the most deformed of the isotopic chain. We obtain rather flat
profiles around sphericity in $^{28,30}$Mg, whose correct
description would need beyond mean field techniques involving
configuration mixing of the shape fluctuations. $^{32}$Mg becomes
also spherical, thus showing that the magic number $N$=20 exists
for the ground-state of $^{32}$Mg in the mean-field theories
\cite{Ren96}. For heavier isotopes the prolate deformation grows
with the increase of the neutron number for $^{32-36}$Mg. While
the prolate solution is only a shoulder in $^{32}$Mg at an energy
higher than that of the spherical ground state, it becomes already
ground state in $^{34}$Mg but practically degenerate with shapes
in the range of $-0.15 < \beta < 0.40$. Finally $^{36}$Mg becomes
again a well deformed nucleus with $\beta \sim 0.3$. This
situation corresponds to the result obtained in
Ref.~\cite{Hamamoto2007} that the weakly-bound neutrons in
$^{33-37}$Mg nuclei may prefer to being deformed due to the
Jahn-Teller effect. In general, we find almost identical values of
the quadrupole parameter $\beta$ with the three Skyrme
parametrizations.

\begin{figure*}
\centering
\includegraphics[width=130mm]{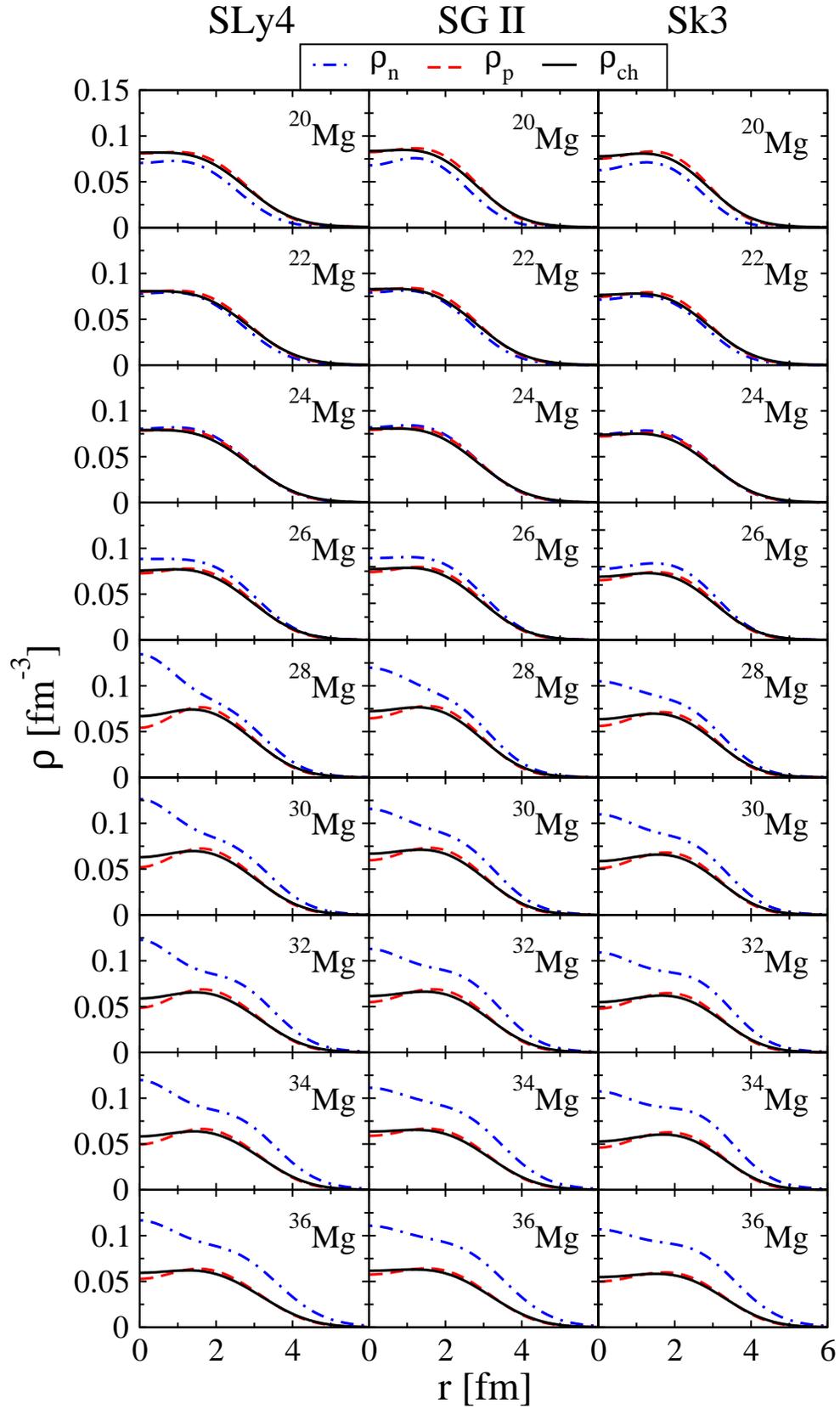}
\caption[]{(Color online) Proton, neutron, and charge density
distributions of even-even Mg isotopes ($A$=20--36) calculated
with SLy4, SGII, and Sk3 interactions.
\label{fig3}}
\end{figure*}

In Fig.~\ref{fig3} we show the proton, neutron, and charge density
distributions of even-even Mg isotopes ($A$=20--36) calculated
with three Skyrme forces. From top to bottom we see the evolution
of these densities as we increase the number of neutrons. Starting
from the lighter isotope $^{20}$Mg we see that the neutron density
is clearly below the proton one, as it corresponds to a larger
number of protons in that isotope ($Z$=12, $N$=8). At the
$N$=$Z$=12 nucleus $^{24}$Mg, we see that both proton and neutron
densities are practically the same except for Coulomb effects that
make the protons to be a little bit more extended spatially, an
effect that has to be compensated with a small depression in the
interior of the proton density. The effect of adding more and more
neutrons is to populate and extend the neutron densities. The
proton distributions try to follow the neutron ones, thus
increasing their spatial extension. This radius enlargement in the
case of protons creates a depression in the nuclear interior to
preserve the normalization to the constant number of protons
$Z$=12. It is also worth noting the sudden increase of the neutron
density at the origin in $^{28}$Mg that corresponds to the filling
of the $s_{1/2}$ orbital. The increase of the density in the
central region leads to a more compressed nucleus, thus having
smaller radius around this isotope.

\begin{figure}
\centering
\includegraphics[width=80mm]{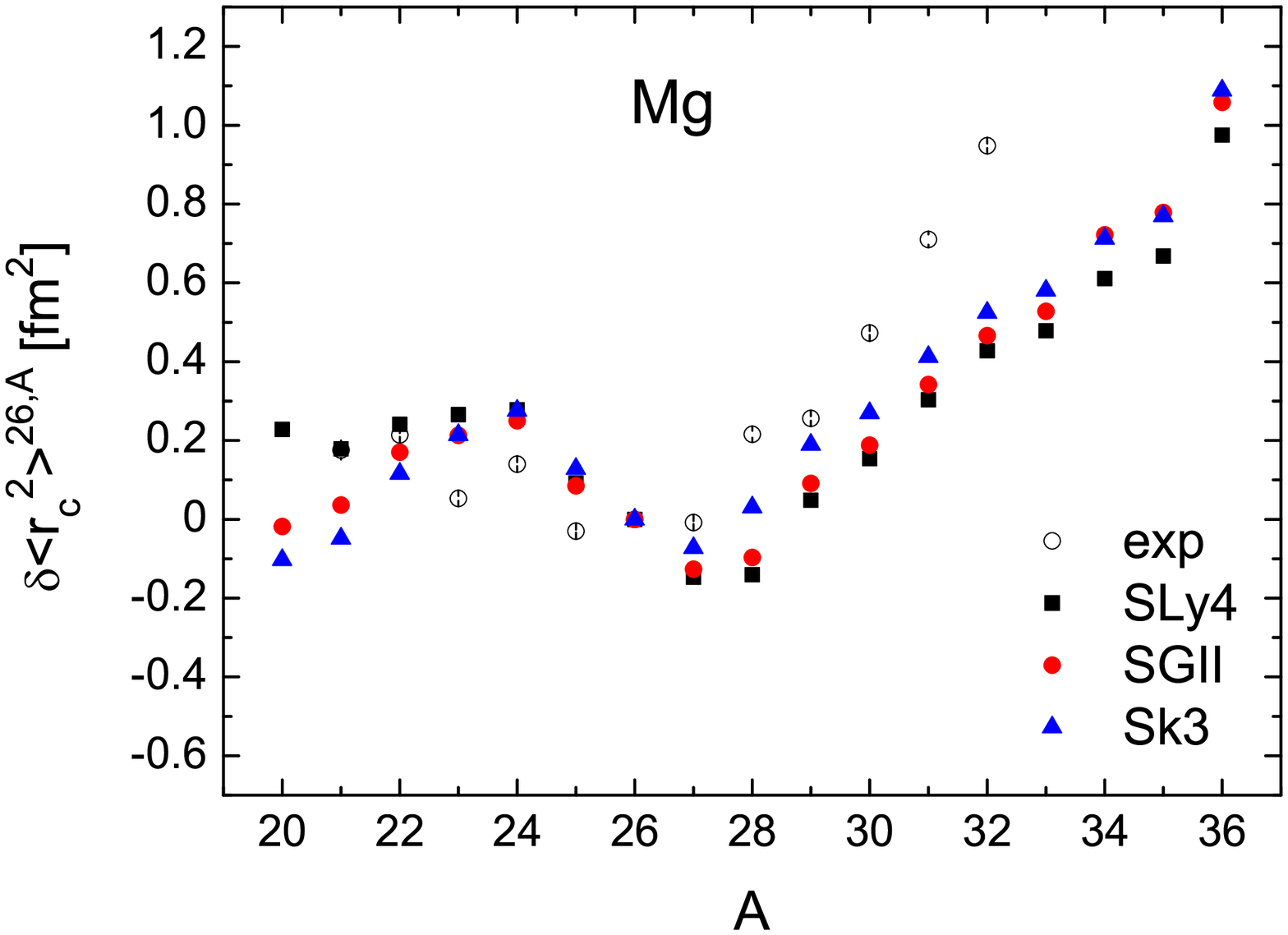}
\caption[]{(Color online) Theoretical (with different Skyrme
forces) and experimental \protect\cite{Yordanov2012} isotope
shifts $\delta<r_{c}^{2}>$ of magnesium isotopes relative to
$^{26}$Mg.
\label{fig4}}
\end{figure}

The charge radius is related to the deformation and the isotope
shifts of charge radii can be used to investigate the deformations
in the isotopic chains. Our results for the squared charge radii
differences $\delta <r^2_{c}>^{26,A}=<r^2_{c}>^A-<r^2_{c}>^{26}$
taking the radius of $^{26}$Mg as the reference are compared in
Fig.~\ref{fig4} with the experimental data \cite{Yordanov2012}. In
general, different Skyrme forces do not differ much in their
predictions of charge rms radii of magnesium spanning the complete
$sd$ shell. The trend of the behavior of the experimental points
and theoretical values strongly corresponds to the neutron shell
structure. For $^{21-26}$Mg isotopes the charge distribution is
compressed due to the filling of the $d_{5/2}$ orbital and the
charge radii do not fluctuate too much. The addition of more
neutrons on either $s_{1/2}$ or $d_{3/2}$ in the range
$^{28-30}$Mg results in a fast increase of the radius. Finally,
for isotopes beyond $^{30}$Mg, where the "island of inversion"
does exist in terms of the rms charge radius \cite{Yordanov2012},
the theoretical results underestimate clearly the experimental
points. Obviously, an additional treatment is needed to understand
in more details this specific region. We note the intermediate
position of $^{27}$Mg, where a minimum is observed in
Fig.~\ref{fig4}, since one of the neutrons added to $^{25}$Mg
fills the last $d_{5/2}$ hole and the other one populates the
$s_{1/2}$ subshell.

\begin{figure}
\centering
\includegraphics[width=80mm]{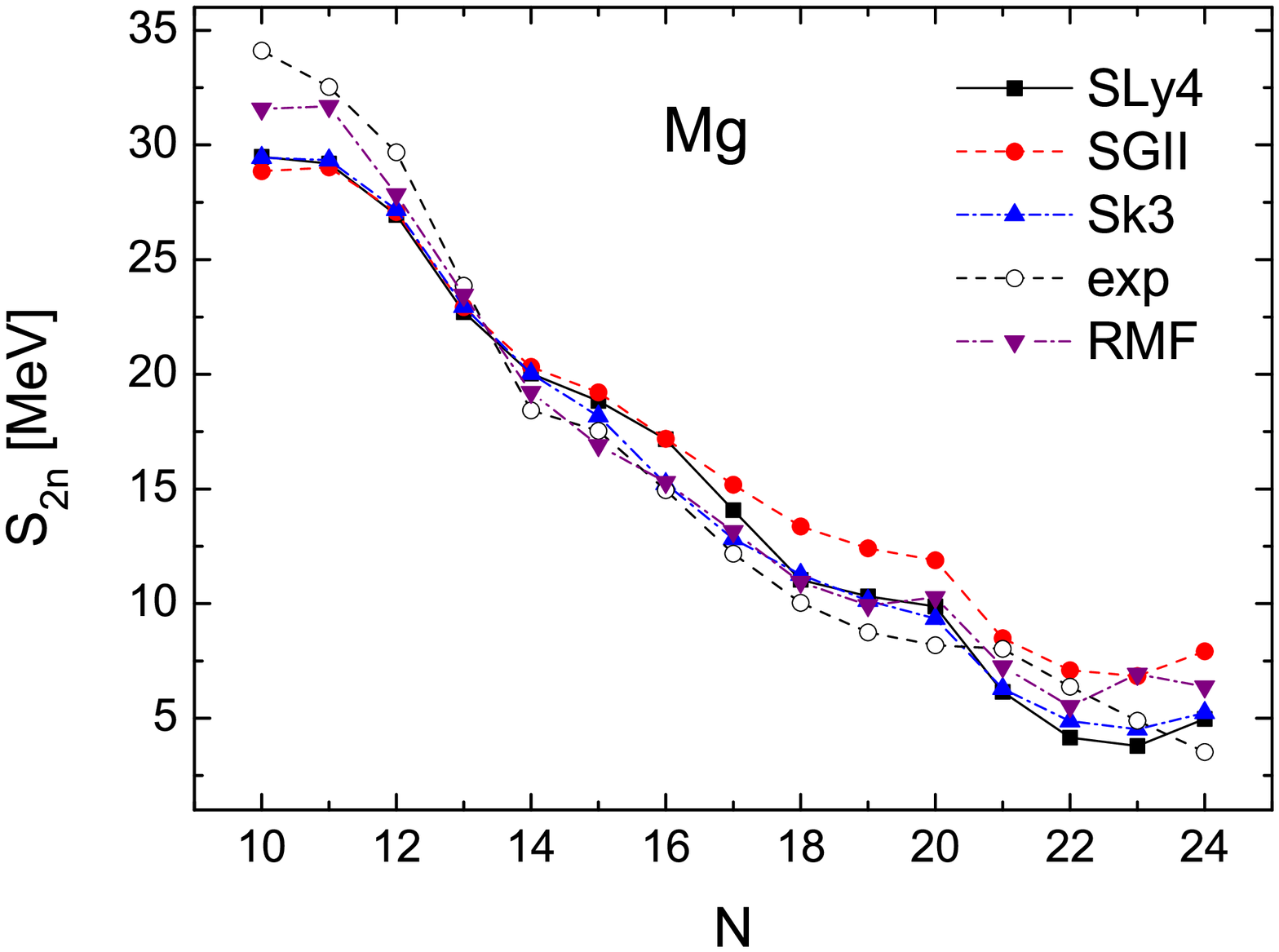}
\caption[]{(Color online) Experimental \protect\cite{Audi2003}
two-neutron separation energies $S_{2n}$ of Mg isotopes for
$A$=20--36, compared to present calculations (with three Skyrme
force parametrizations). Results of RMF calculations are taken
from Ref.~\protect\cite{Ren96}.
\label{fig5}}
\end{figure}

The energy gap between nuclear orbitals and the location of
nuclear shell closures is not static but is subject to the
proton-to-neutron ratio. An example of experimental observable
that probes the location of shell and subshell closures is the
two-neutron separation energy $S_{2n}$. Its trend across isotopic
chains provides a fundamental indication for completely filled
neutron shells. The two-neutron separation energy can be
calculated out of the binding energies of neighbouring even nuclei
$S_{2n}=BE(A,Z)-BE(A-2,Z)$. Results for $S_{2n}$ are plotted on
Fig.~\ref{fig5} together with the experimental data
\cite{Audi2003}. It can be seen that the profile of $S_{2n}$ is
rather similar for all Skyrme parametrizations. The results with
SGII force start to deviate from the others at $N>15$
overestimatig the experiment. The general features of the
experimental data are satisfactorily reproduced. Especially, the
two-neutron separation energy drops at the shell closure $N$=20
because neutrons populating orbitals outside closed shells are
less bound. This is related to the fact that neutrons in a
(nearly) closed neutron shell are more strongly bound and more
energy is needed to remove them out of the nuclear medium. Our
results for $S_{2n}$ are also in a good agreement with the
predictions of the RMF theory with force parameters NL-SH
\cite{Ren96}.

\begin{figure}
\centering
\includegraphics[width=80mm]{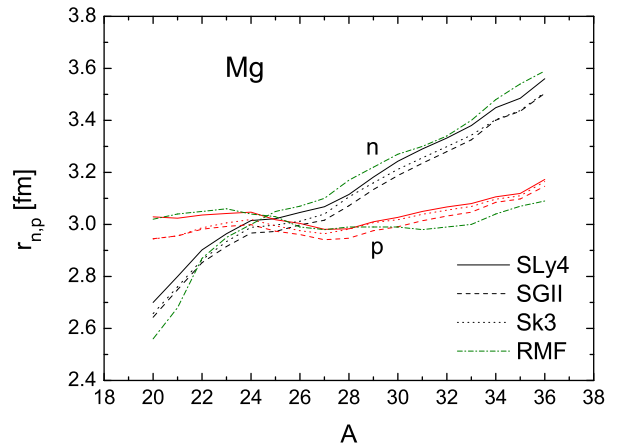}
\caption[]{(Color online) Proton $r_{p}$ (red curves) and neutron
$r_{n}$ (black curves) rms radii of Mg isotopes calculated by
using SLy4, SGII, and Sk3 forces. The results from RMF
calculations \protect\cite{Ren96} (green curves) are also given.
\label{fig6}}
\end{figure}

\begin{figure}
\centering
\includegraphics[width=80mm]{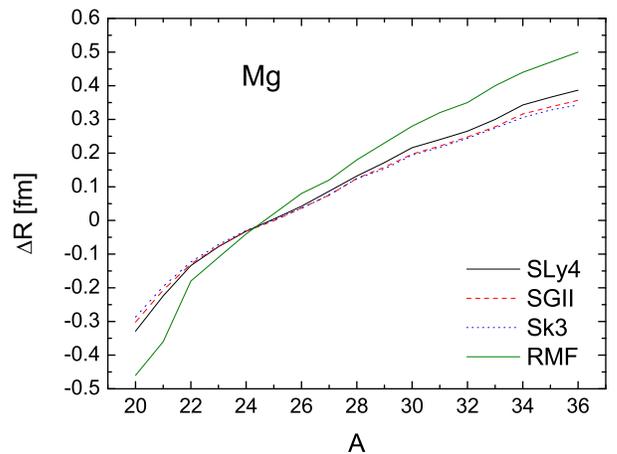}
\caption[]{(Color online) Difference between neutron and proton
rms radii $\Delta r_{np}$ of Mg isotopes calculated by using SLy4,
SGII, and Sk3 forces. The RMF calculation results are from
Ref.~\protect\cite{Ren96}.
\label{fig7}}
\end{figure}

Figure~\ref{fig6} contains our results with the SLy4, SGII, and
Sk3 forces for the neutron and proton mean square radii in the Mg
isotopic chain. They are compared with the predictions from RMF
theory \cite{Ren96}. We see that the tendency in the radii as a
function of the mass number $A$ is quite similar in both
approaches, but the proton rms radii with Skyrme are larger than
the results from RMF for isotopes heavier than $^{28}$Mg. In the
case of neutrons, RMF radii are larger than the Skyrme ones for
isotopes heavier than $^{24}$Mg. As a result we will get
systematically differences between the neutron and proton rms
radii of isotopes with $A>24$, which are larger in the case of RMF
as compared to the case of Skyrme forces. This is clearly seen in
Fig.~\ref{fig7} where we plot the differences between the rms
radii of neutrons and protons $\Delta R=r_n-r_p$. The latter is a
simple measure of a neutron (proton) skin emergence in Mg isotopes
from the considered isotopic chain. We can see from
Fig.~\ref{fig7} that $\Delta R$ increases monotonically with
neutron excess in the chain of Mg isotopes. Moreover, an
irregularity does not seem to be present around $^{32}$Mg neither
in the trend of the neutron and proton rms radii (see
Fig.~\ref{fig6}) nor for their difference. From the Skyrme HF
analysis of the rms radii of proton and neutron density
distributions Lenske {\it et al.} \cite{Lenske98} have found a
proton skin in the neutron-deficient Mg isotopes, while at the
neutron dripline the Mg isotopes develop extended neutron skins.
It can be also seen from Fig.~\ref{fig7} that, in particular when
using SLy4 force, a rather pronounced proton skin in $^{20}$Mg
develops which exceeds the neutron density by 0.33 fm. The latter
value almost coincides with the value of 0.34 fm that has been
obtained in Ref.~\cite{Lenske98} using a standard Skyrme
interaction \cite{Friedrich86}. In the neutron-rich Mg isotopes
our calculations predict quite massive neutron skins whose
thickness for $^{36}$Mg nucleus reaches 0.39 fm when the same SLy4
parametrization is used. It is known that relativistic models
predict rather large neutron radii compared with the
nonrelativistic ones because the saturation density of asymmetric
matter is lower in the EOS when phenomenological nucleon
interaction in the RMF theory is used \cite{Typel2001,Oyamatsu98}.
The results shown for neutron radii of Mg isotopes in
Fig.~\ref{fig6} and correspondingly for neutron and proton
thicknesses in Fig.~\ref{fig7} are consistent with the above
general conclusion. The same tendency has been observed in
Ref.~\cite{Sarriguren2007} for Sn, Ni, and Kr isotopes, for which
different definitions for the skin thickness were tested in the
framework of the same deformed Skyrme HF+BCS model.

\begin{figure}
\centering
\includegraphics[width=80mm]{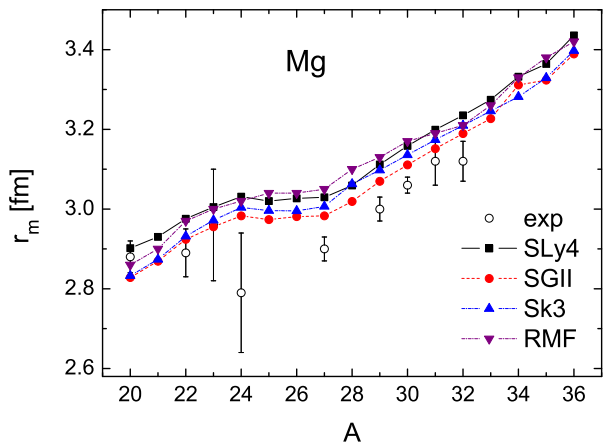}
\caption[]{(Color online) Matter $r_{m}$ rms radii of Mg isotopes
($A$=20--36) calculated by using SLy4, SGII, and Sk3 forces. The
experimental values are taken from Ref.~\protect\cite{Suzuki98}.
The results from RMF calculations \protect\cite{Ren96} are also
given.
\label{fig8}}
\end{figure}

A comparison between the matter rms radii $r_{m}$
[Eq.~(\ref{rmsrm})] obtained from the Skyrme HF+BCS calculations
and their values deduced from the measured interaction cross
sections $\sigma_{I}$ using a Glauber-type calculation
\cite{Suzuki98} is presented in Fig.~\ref{fig8}. The latter were
derived by analyzing the data for $\sigma_{I}$ for Mg isotopes
with use of the Fermi-type distribution under extreme assumptions.
Nevertheless, a reasonable agreement is achieved between both
theoretical and experimental results for the matter rms radii, as
it is seen from the Figure. In addition, the RMF predictions for
$r_{m}$ that are also shown in Fig.~\ref{fig8} essentially agree
with our results and the data. In general, the growing of the
calculated matter rms radii of Mg isotopes follows the same trend
slowly overestimating the experimental data. Due to the large
uncertainty of the $\sigma_{I}$ result obtained for the stable
$^{24}$Mg isotope \cite{Suzuki98}, a relatively small value is
deduced for its matter radius. The systematic measurements of
$\sigma_{I}$, as well as of the total reaction cross sections
$\sigma_{R}$, of Mg isotopes on a proton or complex target (e.g.,
Ref.~\cite{Horiuchi2012} for Mg+$^{12}$C cross section data) may
lead to additional information about nuclear deformation through
the enhancement of the nuclear size.

\begin{figure}
\centering
\includegraphics[width=80mm]{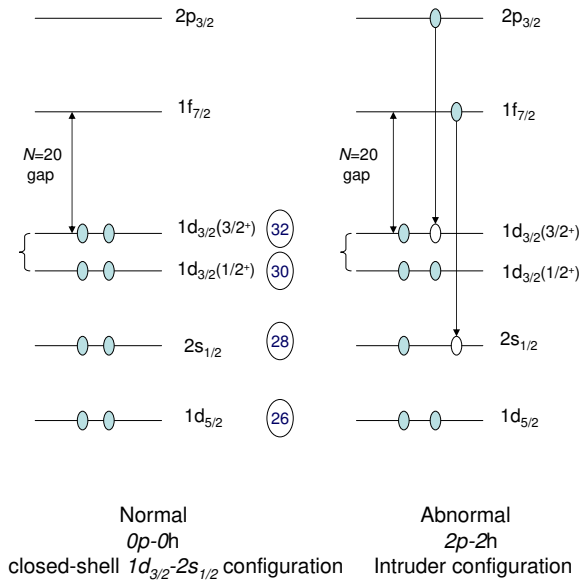}
\caption[]{(Color online) The closed-shell and intruder
configurations for $^{32}$Mg nucleus.
\label{fig9}}
\end{figure}

Along this line, to understand better the specific neutron
shell-model structure leading to a concept of an "island of
inversion" two configurations for $^{32}$Mg are displayed in
Fig.~\ref{fig9}: the closed-shell configuration and the one
consisting of two neutrons excited from the $1d_{3/2}$ and
$2s_{1/2}$ orbitals into the $1f_{7/2}$ and $2p_{3/2}$ orbitals
across the $N=20$ shell gap, making a two-particle, two-hole
state. It is well presumed that namely this promotion of a neutron
pair results in deformed $2p$-$2h$ intruder ground state from the
$fp$ shell which competes with the excited (at 1.06 MeV) spherical
normal neutron $0p$-$0h$ state of the $sd$ shell.

\begin{figure}
\centering
\includegraphics[width=65mm]{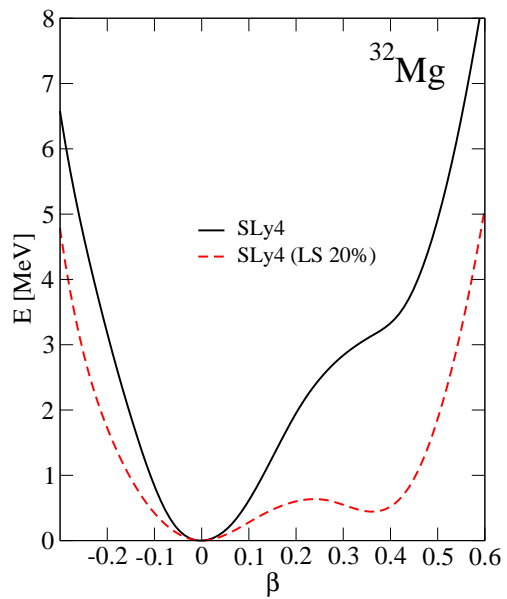}
\caption[]{(Color online) Potential energy curves of $^{32}$Mg
obtained from HF+BCS calculations with SLy4 force for the
spherical case (black solid line) and in the case when the
spin-orbit strength of the effective SLy4 interaction is increased
by 20\% (red dashed line).
\label{fig10}}
\end{figure}

As is well known, the spin-orbit interaction and the pairing
correlations have influence on the deformation of nuclei.
Therefore, we perform additional calculations for the $^{32}$Mg
nucleus by increasing the spin-orbit strength of the SLy4
effective interaction by 20\%. Increasing the spin-orbit strength
will bring near the neutron $f_{7/2}$ and $d_{3/2}$ orbitals
facilitating the promotion of neutrons to the former. According to
the Federman and Pittel mechanism \cite{FedermanPittel79} protons
in the open $d_{5/2}$ orbital overlap substantially with $f_{7/2}$
neutrons ($\ell_p$=$\ell_n$-1) generating nuclear deformation by
the effect of the isoscalar part of the {\it n}-{\it p}
interaction. The corresponding potential energy curve is
illustrated in Fig.~\ref{fig10} together with the curve from the
original SLy4 interaction leading to a spherical equilibrium shape
in $^{32}$Mg. As a result, we find strong prolate deformation for
the intruder configuration ($\beta$=0.38). This value of the
quadrupole deformation is close to the value $\beta$=0.32 found
for the generator coordinate in Ref.~\cite{Yordanov2012}, where a
slight modification of the spin-orbit strength of the effective
interaction for a better description of the "island of inversion"
was also applied. Similar effect of the pairing strength on the
deformation of $^{32}$Mg has been observed in \cite{Ren96}, namely
that stronger pairing force of neutrons and weaker pairing force
of protons lead to a larger deformation of the ground state for
$^{32}$Mg. Thus, the "dual" nature of the latter that is shown in
Fig.~\ref{fig9} and that reflects the shape coexistence in
$^{32}$Mg is in favor to understand the structure of $^{32}$Mg.

\begin{figure}
\centering
\includegraphics[width=80mm]{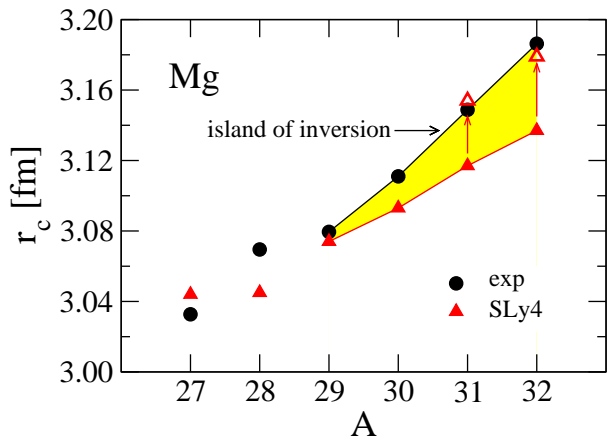}
\caption[]{(Color online) Theoretical (with the SLy4 Skyrme force)
and experimental \protect\cite{Yordanov2012} rms charge radii
$r_{c}$ of Mg isotopes in the range $A$=27--32. The open red
triangles represent the calculated values of $r_{c}$ when the
spin-orbit strength of the effective interaction is increased by
20\%.
\label{fig11}}
\end{figure}

The impact of these new modified calculations on the evolution of
the charge radii, especially in the region of the Mg isotopic
chain where an "island of inversion" is expected, is illustrated
in Fig.~\ref{fig11}. In addition to $^{32}$Mg, we apply the same
procedure also to $^{31}$Mg nucleus in order to establish better
the border of the island. A further increase of the charge radii
of these isotopes is found. For $^{31}$Mg the charge rms radius
increases from 3.117 fm to 3.154 fm and for $^{32}$Mg from 3.137
fm to 3.179 fm toward the experimentally extracted values for both
nuclei indicated in Fig.~\ref{fig11}.  In general, it can be seen
from Fig.~\ref{fig11} that the comparison between the new values
that are very close to the experimental data \cite{Yordanov2012}
and the previously obtained values of the charge radii of
$^{31,32}$Mg isotopes can define a region associated with the
"island of inversion" which is not seen in the HF+BCS theoretical
method by using the original Skyrme force fitted to stable nuclei.

\subsection{Nuclear symmetry energy and its density dependence}

In this subsection we present our results for the symmetry energy
$s$ and the neutron pressure $p_{0}$ in Mg isotopes from the
considered isotopic chain applying the CDFM scheme (see
Section~II), as well as their relationship with the neutron skin
thickness $\Delta R $. The symmetry energy and the pressure are
calculated within the CDFM according to Eqs.~(\ref{eq:10}) and
(\ref{eq:11}) by using the weight functions (\ref{eq:9})
calculated from the self-consistent densities in
Eq.~(\ref{eq:21}). The differences between the neutron and proton
rms radii of these isotopes [Eq.~(\ref{eq:28})] are obtained from
HF+BCS calculations using three different Skyrme forces, SLy4,
SGII, and Sk3.

\begin{figure}
\centering
\includegraphics[width=75mm]{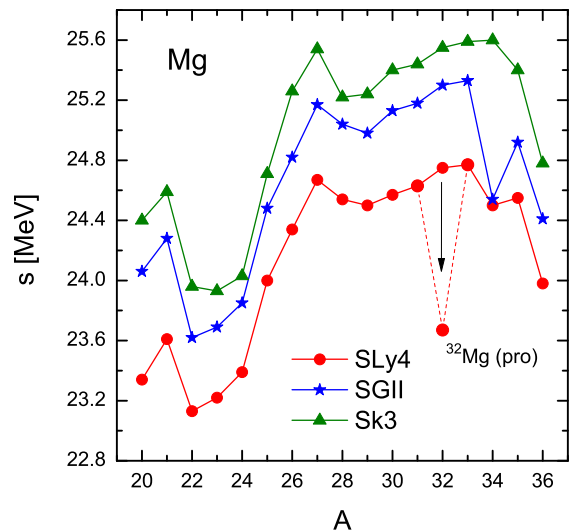}
\caption[]{(Color online) The symmetry energies $s$ for Mg
isotopes ($A$=20--36) calculated with SLy4, SGII, and Sk3 forces.
\label{fig12}}
\end{figure}

The results for the symmetry energy $s$ [Eq.~(\ref{eq:10})] as a
function of the mass number $A$ for the whole Mg isotopic chain
($A$=20--36) are presented in Fig.~\ref{fig12}. It is seen that
the SGII and Sk3 forces yield values of $s$ comparable with each
other that lie above the corresponding symmetry energy values when
using SLy4 set. Although the values of $s$ slightly vary within
the Mg isotopic chain (23--26 MeV) when using different Skyrme
forces, the curves presented in Fig.~\ref{fig12} exhibit the same
trend. It is useful to search for possible indications of an
"island of inversion" around $N$=20 revealed also by the symmetry
energy. Therefore, it is interesting to see how the trend of the
symmetry energy will be changed when for the magic number $N$=20 a
prolate deformed ground state of $^{32}$Mg is obtained (see
Fig.~\ref{fig10}). We would like to note that, in this case, the
modification of the spin-orbit strength of the SLy4 effective
interaction by increasing it by 20\% leads to a smaller value of
$s$=23.67 MeV compared with the one for the spherical case
$s$=24.75 MeV. Thus, the role of deformation on the nuclear charge
radii is also confirmed on the nuclear symmetry energy. Indeed,
the results shown in Fig.~\ref{fig12} are related to the evolution
of the quadrupole parameter $\beta $ as a function of the mass
number $A$ that is presented in Fig.~\ref{fig2}, as well as to the
evolution of the charge radii in Fig.~\ref{fig4}. We find strong
deformations in the range $A$=22-24 that produce larger charge
radii in relation to their neighbors, and local wells in the
symmetry energy. Next, there is a region of flat energy profiles
that correspond to small charge radii and increasing values of the
symmetry energy. Above this region we find first spherical shapes
that produce a plateau in the symmetry energy from $A$=28 to
$A$=34 and finally prolate deformations in $A$=36 that produce
very large radii and a sharp decrease in the symmetry energy. This
confirms the physical interpretation given in
Ref.~\cite{Horowitz2014}, where this fact is shown to be a result
from the moving of the extra neutrons to the surface thus
increasing the surface tension but reducing the symmetry energy.
Although the considered Mg chain does not contain a double-magic
isotope, it is worth mentioning that we find maximum values within
a plateau around the semi-magic isotope $A$=32 that resembles the
sharp peak observed in previous works including double-magic
nuclei \cite{Gaidarov2011,Gaidarov2012}.

\begin{figure}
\centering
\includegraphics[width=80mm]{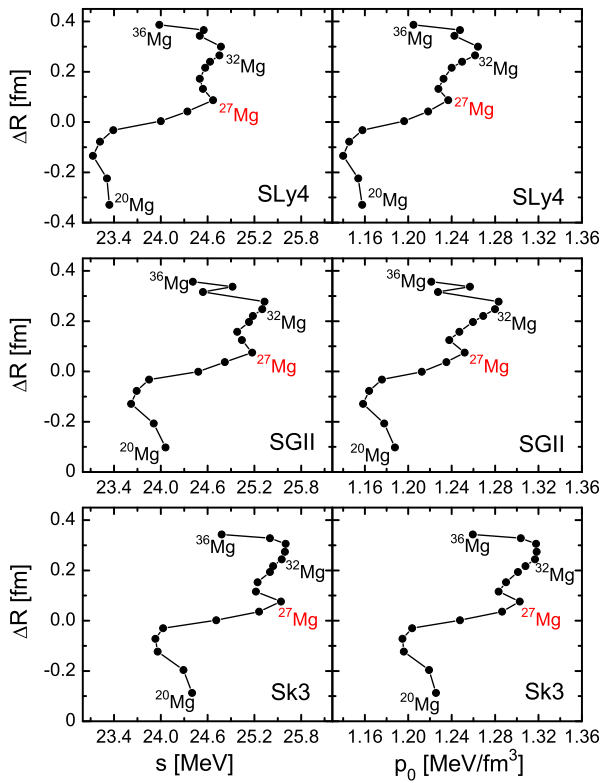}
\caption[]{(Color online) HF + BCS neutron skin thicknesses
$\Delta R$ for Mg isotopes as a function of the symmetry energy
$s$ and the pressure $p_{0}$ calculated with SLy4, SGII, and Sk3
forces.
\label{fig13}}
\end{figure}

We show in Fig.~\ref{fig13} the correlation of the neutron-skin
thickness $\Delta R$ [Eq.~(\ref{eq:28})] of Mg isotopes with the
$s$ and $p_{0}$ parameters extracted from the density dependence
of the symmetry energy around the saturation density. It can be
seen from Fig.~\ref{fig13} that, in contrast to the results
obtained in Refs.~\cite{Gaidarov2011, Gaidarov2012}, there is no
linear correlation observed for the Mg isotopic chain. This
behaviour is valid for the three Skyrme parametrizations used in
the calculations. Such a non-linear correlation of $s$ and $p_{0}$
with the neutron skin thickness $\Delta R$ can be explained with
the fact that stability patterns are quite irregular within this
Mg isotopic chain, where anomalies in shell closures around $N$=20
leading to increased quadrupole collectivity exist. Additionally,
we find the same peculiarity at $A$=27 from Fig.~\ref{fig4}
exhibited in the case of the charge rms radii just reflecting the
transition regions between different nuclear shapes of Mg isotopes
in the considered chain and a small change in the behaviour for
nuclei heavier than $^{32}$Mg, as well.

\section{Conclusions}

In this work, we study nuclear properties of Mg isotopes by means
of a theoretical approach to the nuclear many-body problem that
combines the coherent density fluctuation model \cite{Ant80,AHP}
and the deformed HF+BCS method (with Skyrme-type density-dependent
effective interactions \cite{vautherin}). Three Skyrme
parametrizations were involved in the calculations: SLy4, SGII,
and Sk3. The CDFM makes the transition from the properties of
nuclear matter to the properties of finite nuclei allowing to
investigate the nuclear symmetry energy $s$ and the neutron
pressure $p_{0}$ in finite nuclei on the base of the Brueckner
energy-density functional for infinite nuclear matter. The
isotopes investigated in this work go from the proton-drip-line
nucleus $^{20}$Mg up to $^{36}$Mg that approaches the
neutron-drip-line.

The deformation energy curves show several transitions of
equilibrium shapes as more and more neutrons are added. The lowest
isotope $^{20}$Mg has an energy minimum at zero deformation
parameter value, corresponding to the expected equilibrium
spherical shape for a semi-magic ($N$=8) nucleus. The next two
isotopes, $^{22}$Mg and $^{24}$Mg, have well defined prolate
equilibrium shapes with deformation parameter values around 0.4,
while the next one, $^{26}$Mg, shows shape coexistence with oblate
and prolate equilibrium shapes very close in energy (the oblate
minimum being somewhat deeper than the prolate one). As two or
four  neutrons are added the nucleus becomes soft, with a flat
minimum in an extended region around $\beta$=0, till we reach the
next semi-magic nucleus $^{32}$Mg, which again has spherical
equilibrium shape. However, in this new semi-magic nucleus one can
appreciate a tendency for the nucleus to become prolate, as
indicated by the shoulder on the right hand side of the energy
profile. Finally, the nucleus changes again to soft or prolate
when adding two ($^{34}$Mg) or four ($^{36}$Mg) more neutrons,
respectively. As the proton number is below half-filled-$sd$
shell, the well defined equilibrium prolate shapes take place when
the neutron numbers are either below half-$sd$-shell, or below
half-$fp$-shell. The neutron density profiles are underneath the
proton ones when $N<Z$, practically overlapping when $N$=$Z$, and
exceed them when $N>Z$, as expected, showing a sudden strong
increase in the central region at $N$=16. The central bump that
appears at this $N$ value is due to the occupancy of the $s_{1/2}$
shell and remains for larger $N$ values. On the other hand, the
proton density profiles tend to develop a central hole with
increasing values of $A$ beyond $A$=28 in order to maintain the
proton surface as close as possible to the neutron one. The charge
and mass radii follow the trends observed in the experiment, with
fluctuating values up to $A$=26, and smoothly increasing values
with $A$, beyond $A$=27. These global properties are found to be
rather similar with the three Skyrme forces.

The observed tendency for $^{32}$Mg to become deformed has been
confirmed by repeating our calculations with SLy4 force modifying
slightly the spin-orbit interaction. An increase by a 20\% of the
spin-orbit strength is sufficient to transform the above mentioned
shoulder of the energy profile into a minimum at $\beta$=0.38,
reflecting the role of the intruder $fp$-shell, mainly $f_{7/2}$
intruder. The charge radii calculated at the deformed minima
obtained with the increased spin-orbit interaction for $A$=31, 32
are larger (the increase is from 3.117 fm to 3.154 fm in $^{31}$Mg
and from 3.137 fm to 3.179 fm in $^{32}$Mg), and  are in very good
agreement with experiment. Similar effects are caused by slightly
changing the proton and neutron pairing strengths. These findings
are consistent with results of other theoretical calculations
(e.g.,  within the generator coordinate method) and support the
interpretation of this nuclear region as an "island of inversion".

The correlations of the the neutron skin thickness $\Delta R$ with
the symmetry energy $s$ and with the neutron pressure $p_{0}$ do
not exhibit linear behavior. They show up the same peculiarities
at $A$=27 that reflect the transition regions between different
nuclear shapes of Mg isotopes in the considered chain discussed
above. The values of the symmetry energy $s$ vary roughly between
23 and 26 MeV, being larger for Sk3 force, smaller for SLy4 force
and in between for the SGII effective interaction. Even more
dramatic is the considerable change in the trend of symmetry
energy evolution with the mass number, when we include the results
for the prolate ($\beta$=0.38) ground-state of $^{32}$Mg obtained
with the spin-orbit modified SLy4 effective interaction. The
behavior of these correlations in this isotopic chain is quite
different from that found in our former studies on very heavy
isotopic chains, where we found smoother patterns and more
regularities. This is clearly due to the fact that shell effects
are much more pronounced in these lighter isotopic chains than
they are in the heavy ones.

To probe nuclear structure models it is necessary to have
unambiguous experimental observables for the location of shell and
subshell closures. Therefore, the study of the nuclear level
inversion and nuclear bubble phenomenon would be more complete
after performing electron scattering off short-lived nuclei on the
new generation electron-nucleus RIB facilities. Concluding, we
would like to note that further study is necessary to prove
theoretically the existence of an "island of inversion" probed by
the REX-ISOLDE experiment. In particular, it is worth to perform
calculations by including effects of tensor and three-body forces
and exploring novel energy density functionals.

\begin{acknowledgments}
Two of the authors (E.M.G. and P.S.) acknowledge support from
MINECO (Spain) under Contracts FIS2011--23565 and FPA2010--17142
and from Unidad Asociada I+D+i between IEM-CSIC and Grupo de
F\'\i sica Nuclear (UCM).

\end{acknowledgments}

\end{document}